\documentstyle[12pt]{article}
\font\titlefont=cmbx10 scaled \magstep2
\def\Journal#1#2#3#4{{#1} {\bf #2}, #3 (#4)}

\def\LNC{\em Lett. al Nuovo Cimento}

\def\PLB{{\em Phys. Lett.}  B}

\def\PRD{{\em Phys. Rev.} D}

\def\be{\begin{equation}}
\def\ee{\end{equation}}
\def\bea{\begin{eqnarray}}
\def\eea{\end{eqnarray}}

\begin{document}

\begin{center} {\titlefont The Case for Discrete Energy Levels
of a Black Hole }
 
\vskip .3in
JACOB D. BEKENSTEIN\footnote{email:
bekenste@vms.huji.ac.il} \\
\vskip .1in
Racah Institute of Physics, The Hebrew University of
Jerusalem\\  Givat Ram, Jerusalem 91904 ISRAEL
\end{center}

\vskip .2in
\begin{abstract}The adiabatic invariant nature of black hole
horizon area in classical gravity suggests that in quantum
theory the corresponding operator has a discrete spectrum.  I
here develop further an algebraic approach to black hole
quantization which starts from very elementary assumptions,
and proceeds by exploiting symmetry.  It predicts a uniformly
spaced area spectrum for all charges and angular momenta. 
Area eigenvalues are degenerate;  correspondence with black
hole entropy then dictates a precise value for the interval
between eigenvalues.
\end{abstract}
\vskip .1in

\baselineskip=14pt

\section{Introduction}
\label{I}

To many the term ``black hole'' means a macroscopically large
object, the engine behind the flickering X-ray sources in the
Galaxy, the quasars and other active galactic nuclei.  Talking
about the discrete energy spectrum of a black hole, a patently
quantum issue, must seem incredibly pedantic: why not just use
classical physics ?  But, of course, quantum dynamics is the
underlying dynamics of the world, and the simplicity of black
holes just begs us to use them to learn about quantum
gravity.  The nature of the energy spectrum of a black hole is
about the simplest question that can be asked in this context. 
Here I would like to approach the issue from a unconventional
angle.

One usually discusses an energy spectrum for a collection of
stationary states.  Classically stationary states of a black
hole are described by the Kerr-Newman (KN) joint solution  to
the Einstein and Maxwell equations~\cite{MTW}.  Its
parameters, mass $M$, electric charge $Q$ and angular momentum
${\bf J}$, can all be inferred from the asymptotic behavior of
the electromagnetic field and geometry, and are thus
observables in a true sense.  The  special case ${\bf J}=0$ of
KN is called the Reissner-Nordstr\" om black hole; that with
${\bf J}=0$ and $Q=0$ corresponds to the original Schwarzschild
black hole.  For a quarter century it was widely accepted that
the KN black holes account for all stationary black states
within general relativity and a number of similar gravity
theories, at least if one overlooks trivial extensions of KN
like including Dirac monopole number alongside electric
charge.  This ``no hair principle'' was overturned in the
1990's with the appearance of a number of new black hole
parameters and properties: skyrmion number, nonabelian magnetic
monopole, color, etc.  These complications are irrelevant to
our task.  I will here pretend that $M$, $Q$ and ${\bf J}$ are
the only parameters of stationary black hole states.

\section{Adiabatic Invariance and Black Hole Mass Quantization}
\label{Ad} 

Like any observable of a KN black hole, the area of its event
horizon (boundary)  can be expressed as a function of $M, Q$
and ${\bf J}$:
\be {A=4\pi[ (M+ \sqrt{M^2-Q^2-{\bf J}^2/M^2}\ )^2+{\bf
J}^2/M^2]},
\label{area}
\ee  where I am using units with $G=c=1$ in which $\surd\hbar$
stands for either the Planck's length or the Planck mass.  Any
departure from stationarity is likely to involve an increase in
horizon area~\cite{hawking_area}.  But when {\it slow\/}
changes of a black hole occur, $A$ behaves as an adiabatic
invariant~\cite{bek99,mayo98}.  Recall that an adiabatic
invariant is a quantity which changes especially slowly in the
wake of a slow change of the system's parameter. For an
harmonic oscillator whose cord is gradually extended on a
timescale well exceeding its period, the ratio of the
oscillator's energy to its frequency is constant to exponential
order, and hence an adiabatic invariant.

Lower slowly (adiabatically)  a point charge
$\varepsilon$ down to the horizon of a Reissner-Nordstr\" om
black hole of mass $M$ and charge $Q\gg \varepsilon$, and then
allow it to be assimilated by the hole.  The black hole
obviously experiences changes: $\delta Q=\varepsilon$ and 
$\delta M =\varepsilon\Phi$, where $\Phi$ is the electrostatic
potential at the horizon  (the particle's rest energy is
entirely redshifted away).  From the KN solution we have
\be
\Phi={Q\over M+ (M^2-Q^2)^{1/2}}
\label{phi}
\ee The increment of the horizon's area, Eq.~(\ref{area}), is
\be
\delta A=(\delta M-\Phi\delta Q)\Theta^{-1};\quad \Theta
\equiv  {\scriptstyle 1\over \scriptstyle 2} (M^2-Q^2)^{1/2}
A^{-1}
\label{delta_A}
\ee which evidently vanishes for $\delta Q=\varepsilon$ and 
$\delta M = \varepsilon\Phi$.  Thus to
$O(\delta Q)$ the area is unchanged: classically $A$ and all
smooth functions of it are adiabatic invariants.  This example
can be extended to the Schwarzschild and generic Kerr-Newman
black holes~\cite{bek73}.  Several other examples 
demonstrating adiabatic invariance of horizon area under
electromagnetic or scalar field perturbations have been
presented by Mayo~\cite{mayo98}. The adiabatic invariant nature
of $A$ is useful in at least one astrophysical
context~\cite{duez}, and seems to extend to quantum
gravity~\cite{bar_das_kunst}. 

The basic conclusion of our example is unchanged when the
quantum mechanics of the charge is taken into account.  A
simple calculation~\cite{erice} shows that although there is
now a change in horizon area, $\delta A\geq \xi \hbar$ with
$\xi=O(1)$ and independent of the particle's charge and mass,
and of type and scale of black hole, the minimum area increase
is, again, of second order, and thus negligible.  The
adiabatic invariant status of $A$ is sustained.

Ehrenfest's principle~\cite{born} of the old quantum
mechanics, {\it a classical adiabatic invariant corresponds to
a quantum observable with a discrete spectrum\/}, now implies
that the operator $\hat A$ in the quantum theory  (we denote
operators by carets), which corresponds to classical horizon
area, is expected to have a discrete spectrum:  $\{ a_1, a_2, 
\cdots a_n\cdots \}$.  We may even guess from the fact that
introducing a quantum particle into a KN black hole carries a
minimal ``cost''  $\delta A=\xi\hbar$ which does not depend on
how big the black hole is, that the spacing between area
eigenvalues of the KN area spectrum is a uniform one.  This
way the smallest jump in area corresponds to a transition
between neighboring ``area levels''.

Such a simple structure of the spectrum of area warrants
considering $\hat A$ alongside the charge observable $\hat Q$
and the angular momentum one $\hat {\bf J}$ as the fundamental
observables of a black hole.  From this point of view black
hole mass $\hat M$ is a secondary observable.  We may guess it
is given by the Christodoulou-Ruffini
formula~\cite{chris-ruf}   (obtained by inverting (\ref{area}))
\be  M = \left[{A\over 16\pi}\left(1+{4\pi Q^2\over
A}\right)^2 + {4\pi {\bf J}^2\over A}\right]^{1/2},
\label{mass}
\ee and replacing $A\rightarrow \hat A$, etc.  This together
with the equispaced area  spectrum implies the mass spectrum 
\be  M\sim \sqrt{\hbar n}\quad n=1, 2, \cdots
\label{root_spectrum}
\ee   for the quantum Schwarzschild black
hole~\cite{bek74,mukh86,bek_mukh95}.  

As we shall see, such mass spectrum is inevitably highly
degenerate.  It is also  strongly  at variance with Hawking's
celebrated semiclassical result for the spectrum radiated by a
black hole---a continuum with roughly Planckian shape.  Our 
prediction tells us that quanta emitted as a result of the
black hole jumping one or a few steps of the
$\surd n$ spectrum should have frequencies which are quite
accurately multiples of a frequency $\sim 1/M$, of order of the
peak frequency in Hawking's spectrum.  What does full blown
quantum gravity have to say about this ?  

As summarized in my 2001 Erice lectures~\cite{erice}, a number
of workers employing canonical quantization of gravity have 
obtained a mass spectrum of this form for the Schwarzschild
black hole, but with no consensus as to the exact numerical
coefficient~\cite{many}.  The story is somewhat similar for
charged and rotating black holes~\cite{charged,bar_das_kunst}. 
Some canonical gravity calculations give a not uniformly
spaced spectrum~\cite{others}.  The same is found by loop
quantum gravity methods~\cite{ashtekar}, but there are two
claims that an equispaced area spectrum is consistent with
loop gravity after all~\cite{alekseev}.  In string theory an
extreme black hole (a BPS state) has horizon area which is
quantized as the square root of a product of integers (the
charges).  This is in the spirit of the finding here, but the
rule does not extend to the nonextreme black holes which is
said to posses a continuum mass spectrum.  

If no errors are involved, then the various approaches must be
elucidating the black hole spectrum at various levels of
accuracy.  For example, it may be that (\ref{root_spectrum})
gives the pristine spectrum before any level broadening by
radiative effects or any degeneracy breaking by perturbations
is effective, while the string result describes the situation
after all perturbations are taken into effect.  It would be
nice to find out for sure, and by the simplest logic, what the
pristine spectrum looks like.   In what follows I describe a
purely algebraic approach to the black hole area spectrum
which, to my mind, clarifies this question.

\section{Black Hole Algebra} 
\label{Alg}

Like any other quantum system, the quantum black hole should
exhibit the usual angular momentum spectrum, Spect$(\hat {\bf
J}^2)=\{j(j+1)\hbar^2|\ j=0,{\scriptstyle 1\over \scriptstyle
2}, 1, \cdots\,\}$, Spect$(\hat J_z)=\{m\hbar|\ m=-j\ ,\
-(j-1),\ \cdots,\  j\}$.   This is known to follow  from the
algebra of angular momentum
\be  {\bf \hat J}\times {\bf \hat J}=\imath\hbar {\bf \hat J}
\ee  which we adopt for the black hole.  We likewise expect on
intuitive grounds that there exist one-black hole states,
denoted by   $|njmqs\rangle $, each of which  can be specified
{\it simultaneously\/} by $j$ and $m$ as well as  the 
eigenvalue of charge $\hat Q$, $\{qe|\ q=0, \pm 1, \pm 2,
\cdots\}$ ($e$ denotes the elementary charge), that of
$\hat A$, $\{a_n\, | n=1, 2, 3,\cdots\ \}$, and an additional
quantum number, $s$, which distinguishes between states with
common sets $\{nqjm\}$;  $s=1, 2,\cdots, g_n$; $g_n$ is the
degeneracy of the said states.   Evidently this requires at the
very minimum that
\be [{\bf \hat J}, \hat Q] = 0;\quad [{\bf \hat J}, \hat A] =
0;  \quad [\hat Q, \hat A] = 0.
\label{3}
\ee  The first two conditions are justified because both
charge and area should be invariant under rotations.  The last
follows if area is gauge invariant (charge is the generator of
a global gauge transformations).  Gour~\cite{gour} has
succeeded in defining an operator whose eigenvalues are $s$;
however, he presupposes that the spectrum of $\hat A$ is
equispaced.  Since showing this is one of our goals we choose
a different approach. Finally, we define the black hole
vacuum  $|{\rm vac}\rangle$ (assumed unique) as the state with
vanishing area, $j=0$ and $ q=0$.   

For every basis state define an operator $\hat R_{njmqs}$ such
that $|njmqs\rangle=\hat R_{njmqs}\,|{\rm vac} \rangle$.   This
leaves a lot of freedom in  $\hat R_{njmqs}$ which we shall
exploit to our convenience forthwith.  To imbue the algebra of
${\bf \hat J}, \hat Q, \hat A, \hat I$ (unity operator) and
$\hat R_{njmqs}$ with some nontrivial content, we shall assume
that it is closed and linear.  Closure is assumed for
simplicity: no new operators arise from commuting those
already present.  Linearity (the commutator of any two
operators in the algebra is a linear combination of some of
the operators in the algebra) is a strong assumption because
one might inquire what singles out
$\hat A$ as opposed to, say, $\surd\hat A$ as the operator in
terms of which the algebra is linear. Our ultimate
justification is that the algebra, as stated  and in no other
form, makes
$\hat A$ an {\it additive\/} quantity for systems of several
black holes~\cite{erice}, just as are $\hat Q$ and ${\hat
J}_z$, and just as intuition demands.       

By definition $\hat Q|njmqs \rangle= qe\, |njmqs\rangle $ so
that 
\be
\exp(\imath\chi
\hat Q)\,  \hat R_{njmqs}{|{\rm vac}\rangle} =
\exp(\imath\chi q e)\, \hat R_{njmqs} {|{\rm vac}\rangle}.
\label{gauge1}
\ee Thus $\exp(\imath\chi \hat Q)$ performs a global gauge
transformation on the basis states.  One thus suspects that
the same operator transforms all the $\hat R$ operators as
\be
\exp(\imath\chi \hat Q)\, \hat R_{njmqs}  \exp(-\imath\chi
\hat Q) =  \exp(\imath\chi q e)\,\hat R_{njmqs}
\ee  because operating with this last on ${|{\rm vac}\rangle}$
reproduces Eq.~(\ref{gauge1}).  To $O(\chi)$ this last equation
gives another basic commutator
\be [\hat Q, \hat R_{njmqs}] = q e\, \hat R_{njmqs}.
\label{charge}
\ee     

Passing on we note that under rotations ${|{\rm vac}\rangle}$
must be invariant but
$\hat R_{njmqs}\, |{\rm vac}\rangle$ must transform like the
corresponding spherical harmonics $Y_{jm}$ (or spinorial
harmonic if $j$ is half-integer).  Thus  $\hat R_{njmqs}$ must
be an irreducible spherical tensor operator of rank
$j$~\cite{merzbacher}.  This entails the commutators
\bea &&[\hat J_z, \hat R_{njmqs}] = m\,\hbar\, \hat  R_{njmqs}
\label{Jz}
\\   &&[\hat J_\pm, \hat R_{njmqs}] = \sqrt{j(j+1)-m(m\pm
1)}\,\hbar\,\hat R_{nj,m\pm 1,qs},
\label{Jpm}
\eea where $J_{\pm}=J_x\pm\imath J_y$. How does $\hat A$
commute with the $\hat R_{njmqs}$ ?  We start with Jacobi's
identity $[\hat A, [\hat V, \hat R_{njmqs}]\,] + [\hat V,
[\hat R_{njmqs}, \hat A]\,] + [\hat R_{njmqs}, [\hat A, \hat
V]\,]=0$ with $ \hat V \rightarrow 
\{\hat J_z,  \hat J_\pm,  \hat Q\}$ to obtain 
\bea && [\hat Q, [\hat A, \hat R_{njmqs}]\,]  =  q e\, [\hat A,
\hat R_{njmqs}]\label{one}
\\   && [\hat J_z, [\hat A, \hat R_{njmqs}]  =  m\,\hbar\,
[\hat A,  \hat R_{njmqs}]\label{two}
\\ &&[\hat J_\pm, [\hat A, \hat R_{njmqs}]\,]  = 
\sqrt{j(j+1)-m(m\pm 1)}\,\hbar\,[\hat A,\hat R_{nj,m\pm 1,qs}].
\label{three}
\eea Note that these commutators and those in (\ref{3}) are
invariant under $\hat A\rightarrow \hat A+{\rm
const.}\times\hat I$.  To eliminate the freedom in $\hat A$ we
demand that $\hat A\,{|{\rm vac}\rangle}=0$.  Further, we note
that the commutators (\ref{one}-\ref{three}) parallel those of
$\hat R_{njmqs}$ alone with ${\hat {\bf J}}$ and $\hat Q$
meaning that $[\hat A,
\hat R_{njmqs}]$ transforms under rotations and gauge
transformations exactly like $\hat R_{njmqs}$. 

In view of this, linearity and closure tell us that ($\hat
J_0\equiv \hat J_z, \hat J_{\pm 1}\equiv \hat J_{\pm}$)  
\be [\hat A,
\hat R_{njmqs}]=\sum_{n's'} h_{ns}^{n's'}\, \hat R_{n'jmqs'}
+\delta_{q}^0\big[\delta_{j}^0\, (C_{ns}\hat I + D_{ns}\hat Q
+ E_{ns}\hat A) + \delta_{j}^1 F_{ns}\hat J_{m}\big]
\label{commuteA}
\ee  Here the matrix $ h_{ns}^{n's'}$ could depend on $q, j$
and
$m$, but the various $C_{ns}, D_{ns}, E_{ns}$ and
$F_{ns}$ are just complex constants.  We have not included a
$\hat R_{njmqs}$  in the r.h.s. with $j$ and $q$ other than
seen in the l.h.s. since such operator would transform under
rotations and gauge transformations differently than the
l.h.s.  For like reason the $\hat I, \hat Q$ and $\hat A$, all
gauge and rotationally invariant, occur in the r.h.s. only in
conjunction with $R_{n000s}$ in the l.h.s., and the (spherical)
components of ${\hat {\bf J}}$, which transform like an
irreducible tensor of rank one, but are invariant under gauge
transformations, occur in the r.h.s. only in conjunction with
$R_{n1m0s}$ in the l.h.s.

When we operate on $|{\rm vac}\rangle$ with
Eq.~(\ref{commuteA}),
 ${\bf \hat J}, \hat Q$ and $\hat A$ kill this state, $\hat I$
preserves it, while the various $\hat R_{njmqs}$ create
one-black hole states.  But one cannot represent the vacuum as
a linear superposition of $|njmqs\rangle $.  Thus we must set
$C_{ns}=0$.  Similarly,
$|njmqs\rangle $ with different $n$ have to be orthogonal
(because $\hat A$ is hermitian), and those with different $s$
(even if $n, j, m$ and $q$ are common) can be made orthogonal
by Schmidt orthogonalization.  Therefore we must have
$h_{ns}^{n's'} = k_{ns}\, \delta_n^{n'} \delta_s^{s'}$.  We
are thus left with
\be [\hat A, \hat R_{njmqs}] =k_{ns}\, \hat R_{njmqs}
+\delta_{q}^0\big[\delta_{j}^0\, (D_{ns}\hat Q + E_{ns}\hat A)
+ \delta_{j}^1 F_{ns}\hat J_{m}\big]
\label{A} 
\ee Operating with this on the vacuum shows that necessarily
$k_{ns}=a_n$.  Let us introduce new black hole creation
operators
\be
\hat {\cal R}_{njmqs}\equiv \hat R_{njmqs} +  (a_n)^{-1}
\delta_{q}^0\big[\delta_{j}^0\, (D_{ns}\hat Q + E_{ns}\hat A)+
\delta_{j}^1 F_{ns}\hat J_m\ \big]
\label{redefinition}
\ee Obviously the algebra of ${\bf \hat J}, \hat Q,
\hat A, \hat I$ and $\hat{\cal  R}_{njmqs}$ is still linear and
closed, and $\hat{\cal  R}_{njmqs}$ creates the same state as
$\hat  R_{njmqs}$.  The commutators (\ref{charge})-(\ref{Jpm})
and (\ref{A}) are replaced by
\bea &&[\hat Q, \hat {\cal R}_{njmqs}] = qe\, \hat {\cal
R}_{njmqs}
\label{QR}
\\ &&[\hat J_z, \hat {\cal R}_{njmqs}] = m\,\hbar\, \hat  {\cal
R}_{njmqs}-(m\hbar/a_n)\delta_{q}^0\delta_{j}^1 F_{ns}\hat J_m
\label{RJz}
\\   &&[\hat J_\pm, \hat {\cal R}_{njmqs}] =
\sqrt{j(j+1)-m(m\pm 1)}\,\hbar\,\hat {\cal R}_{nj,m\pm 1,qs}
\nonumber
\\ &&\qquad\qquad\quad - (\sqrt{2-m(m\pm
1)}\hbar/a_n)\delta_{q}^0\delta_{j}^1 F_{ns} 
\hat J_{m\pm 1}
\label{RJpm}
\\ &&[\hat A, \hat {\cal R}_{njmqs}] = a_n\, \hat {\cal
R}_{njmqs}
\label{AR}
\eea The first three equations include contributions which
vanish due to factors of the form $q\, \delta_q^0$,
$\delta_j^0\,\sqrt{j(j+1)-m(m\pm 1)}$, etc. 

\section{The Area Spectrum}
\label{Spec}
 
Henceforth I use the compact notation $\hat {\cal
R}_\lambda\equiv
\hat {\cal R}_{{n_\lambda}{q_\lambda}
{j_\lambda}{m_\lambda}{s_\lambda}}$,
$ D_\kappa\equiv D_{{n_\kappa}{s_\kappa}}$ etc. Evidently by
closure and linearity
\be [\hat {\cal R}_{\kappa}, \hat {\cal R}_{\lambda}]
=\sum_{\mu}
\epsilon_{\kappa\lambda}^{\mu}\hat {\cal R}_{\mu} +  {\cal 
C}_{\kappa\lambda}\hat I + {\cal  D}_{\kappa\lambda}\hat Q +
\sum_m {\cal  F}_{\kappa\lambda}^m\hat J_{m}+ {\cal 
E}_{\kappa\lambda}\hat A
\label{RR} 
\ee with all structure constants antisymmetric in $\kappa$ and
$\lambda$.  I assume that for fixed $\kappa\neq \lambda$, not
all $\epsilon_{\kappa\lambda}^{\mu}$ vanish (see below). 
Obviously $|\kappa,\lambda\rangle\equiv [\hat {\cal
R}_{\kappa},
\hat {\cal R}_{\lambda}]\, {|{\rm vac}\rangle}$ is a
superposition of  basis states $|nqjms\rangle$, i.e., a
one-black hole state.  What is special about it ?  Commute
(\ref{RR}) with
$\hat A$ to get  $\sum_\mu \epsilon_{\kappa\lambda}^{\mu} a_\mu
\hat{\cal R}_\mu$ in the r.h.s. while the Jacobi identity
together with Eq.~(\ref{AR}) gives $(a_\kappa+a_\lambda)[\hat
{\cal R}_{\kappa}, \hat {\cal R}_{\lambda}] $ in the l.h.s. 
Consistency then demands that
${\cal  C}_{\kappa\lambda}= {\cal  D}_{\kappa\lambda}= {\cal 
F}_{\kappa\lambda}^m= {\cal  E}_{\kappa\lambda}=0$ as well as
$a_\mu=a_\kappa+a_\lambda$ whenever
$\epsilon_{\kappa\lambda}^{\mu}\neq 0$.   Repeat the exercise
with $\hat Q$ replacing $\hat A$ and discover that
$q_\mu=q_\lambda+q_\kappa$ ($q_\mu$ is, of course, a physical
value for a charge) whenever
$\epsilon_{\kappa\lambda}^{\mu}\neq 0$.  Thus
$|\kappa,\lambda\rangle$ is an area eigenstate as well as a
charge eigenstate (it satisfies the charge superselection
rule). Therefore,  {\it sums of $a_n$'s belonging to one-black
hole states are possible quantum numbers of physical one-black
hole states\/}.

The assumption that for fixed $\lambda\neq \kappa$ not all
$\epsilon_{\kappa\lambda}^{\mu}$ vanish means $\hat{\cal
R}_\lambda$ and $\hat{\cal R}_\kappa$ never commute. 
Commutativity would  signify, as in field theory, that
creations of the two black holes are independent processes. 
Since we know from the classical limit that two black holes can
merge into one, this last possibility cannot be a general law,
and noncommutativity is the easiest way to introduce the
possibility of black hole fusion already at the quantum
level.      

Operating on ${|{\rm vac}\rangle}$ with the hermitian conjugate
of Eq.~(\ref{AR}), namely
\be [\hat A, \hat {\cal R}_{\kappa}^\dagger]=-a_\kappa \hat
{\cal R}_{\kappa}^\dagger,
\label{hermit}
\ee we discover that $\hat {\cal R}_{\kappa}^\dagger$ must
anhilate
 ${|{\rm vac}\rangle}$ since $\hat A$ is positive definite:
$\hat {\cal R}_\kappa^\dagger$ has the nature of an anhilation
operator.   Now commute  $\hat{\cal R}_{\lambda}$ with
Eq.~(\ref{hermit}) and simplify the triple commutator by means
of Jacobi's identity to get
\be [A,[\hat{\cal R}_\kappa^\dagger, \hat{\cal
R}_\lambda]\,]=(a_\lambda-a_\kappa)[\hat{\cal
R}_\kappa^\dagger,
\hat{\cal R}_\lambda].
\label{RdaggerR}
\ee It is plain that $|\bar\kappa,\lambda\rangle\equiv
[\hat{\cal R}_\kappa^\dagger, \hat{\cal R}_\lambda]\, {|{\rm
vac}\rangle}$, if it is nonvanishing at all, is a one-black
hole state because it is obtained from a one-black hole state
by attempting to anhilate from it a black hole of different
description from the one it actually contains (
$|\bar\kappa,\lambda\rangle$ is not a linear combination with
$|{\rm vac}\rangle$ because, as we shall see, it has definite
area). Now from Eq.~(\ref{RdaggerR}) it is clear that
$|\bar\kappa,\lambda\rangle$ is an eigenstate of $\hat A$ with
eigenvalue $a_\lambda-a_\kappa$.  Of course this only makes
sense when $a_\kappa<a_\lambda$ because $\hat A$ is positive
definite.  In the opposite case $[\hat{\cal R}_\kappa^\dagger,
\hat{\cal R}_\lambda]$ must anhilate ${|{\rm vac}\rangle}$. 
By replacing $\hat A$ in Eqs.~(\ref{hermit})-(\ref{RdaggerR})
by $\hat Q$ one easily shows that $|\bar\kappa,\lambda\rangle$
has the definite physically acceptable charge
$(q_\lambda-q_\kappa)$.   We have thus found that {\it positive
differences of $a_n$'s belonging to one-black hole states are
possible quantum numbers of physical one-black hole states\/}.

{\it A priori\/} the $n$-th (by magnitude) area eigenvalue for
a one-black hole state with definite $j, m, q$ should depend
on all of these: $a_n=a_n(j,m, q)$. But the $m$ dependence here
is excluded by rotational invariance.  Further, combining
states labelled by $j_\lambda$ and $j_\kappa$  should,
according to quantum mechanics, give a state with a
superposition of angular momenta
$|j_\lambda-j_\kappa|, \cdots\, , j_\lambda+j_\kappa$.  But we
know that $|\kappa,\lambda\rangle$ and
$|\bar\kappa,\lambda\rangle$ have definite areas, so it
seems---and we shall so assume pending rigorous proof---that
$a_n$ cannot depend on $j$.  We are left with the dependence
$a_n=a_n(q)$. And charge conjugation symmetry tells us that
the $a_n(q)$ must all be even. 

Obviously $a_1(0)+a_2(0)$, a possible eigenvalue for zero
charge, cannot fall below $a_3(0)$, the third eigenvalue.  But
we cannot allow $a_1(0)+a_2(0)>a_3(0)$, for in that case
$a_1(0)+a_2(0)-a_3(0)$ would be a possible zero-charge area
eigenvalue falling below the lowest possible one,
$a_1(0)$ ! Thus $a_1(0)+a_2(0)=a_3(0)$.  Similarly one shows
that $a_4(0)=a_3(0)+a_1(0)$, and more generally
$a_{n+1}(0)=a_n(0)+a_1(0)$ for $n\geq 2$.  Further, $a_2(0)\geq
2a_1(0)$ for otherwise $a_2(0)-a_1(0)$, also an acceptable
zero-charge area eigenvalue, would fall below
$a_1(0)$.  But if $a_2(0) > 2a_1(0)$, then
$a_1(0)<a_2(0)-a_1(0)<a_2(0)$  so $a_2(0)-a_1(0)$ would be an
eigenvalue in between the first and  second ones !  Thus
$a_2(0)=2a_1(0)$.  Collecting our results we get for the
zero-charge (Schwarzschild) one-black hole area spectrum
\be {\rm Spect}(\hat A\, |\, q=0)=\{na_1(0)\ |\, n=1, 2,
\cdots\ \}
\label{neutral_spectrum}
\ee This shows the promised uniform spacing.

For $q\neq 0$, $a_1(q)+a_1(0)$ corresponds to a state of
charge $q$, and so it cannot fall below $a_2(q)$.  Were
$a_1(q)+a_1(0)-a_2(q)$ positive, it would be a possible area
eigenvalue for zero charge; however, it evidently falls below
$a_1(0)$ so the mentioned option entails a contradiction. 
Thus $a_1(q)+a_1(0)=a_2(q)$.  Similarly one can show that 
$a_n(q)+a_1(0)=a_{n+1}(q)$  for
$n\geq 2$.  Thus the area spectrum for fixed $q\neq 0$ shows
the same uniform spacing as (\ref{neutral_spectrum}), but
possibly a different lowest eigenvalue.  The question now is,
is
$a_1(q)\neq a_1(0)$ ?   

Were $a_1(q)-a_1(0)$ positive, it would be a possible area
eigenvalue for charge $q$, so that  $a_1(q)-a_1(0)\geq
a_1(q)$.  However, this would imply that $a_1(0)=0$ which is
contrary to the basic assumption.  One option is that
$a_1(q)=a_1(0)$ in which case the charge $q$ black hole
spectrum is identical to that of the neutral one,
(\ref{neutral_spectrum}).  The other is that $a_1(0)-a_1(q)$ is
positive and so is a possible area eigenvalue for charge
$-q$.  But $a_n(q)$ is even so we have
$a_1(0)-a_1(q)\geq a_1(q)$, or $a_1(0)\geq 2a_1(q)$.  Further,
$a_1(q)+a_1(-q)$ is evidently  a possible area eigenvalue for
zero charge.  It follows that  $2a_1(q)\geq a_1(0)$.  This is
consistent with our previous finding only if
$a_1(q)={\scriptscriptstyle 1\over
\scriptscriptstyle2}a_1(0)$, which constitutes the second
option.  

I now show that for $q\geq 2$ the second option is possible
only if already $a_1(1)={\scriptscriptstyle 1\over
\scriptscriptstyle 2}a_1(0)$.  First suppose that for some
$\tilde q\geq 1$, $a_1(\tilde q)\neq a_1(\tilde q+1)$.  Then
by the last paragraph  $|a_1(\tilde q)-a_1(\tilde
q+1)|={\scriptscriptstyle 1\over
\scriptscriptstyle2}a_1(1)$.  On the other hand either
$a_1(\tilde q)-a_1(\tilde q+1)$ stands for a state of type
$|\bar \kappa,\lambda\rangle$ with charge $-1$ or 
$a_1(\tilde q+1)-a_1(\tilde q)$ stands for one of charge $+1$.
In either case ${\scriptscriptstyle 1\over \scriptscriptstyle2}
a_1(1)=|a_1(\tilde q)-a_1(\tilde q+1)|\geq a_1(1)$.  But this
is inconsistent if $a_1(1)=a_1(0)$.  Thus if
$a_1(1)=a_1(0)$, $a_1(q)= a_1(q+1)$ also for all $q\geq 1$ and
the spectrum of $\hat A$ for all $q>1$ is just like that for
$q=0$, namely  (\ref{neutral_spectrum}). 

For the alternative $a_1(1)={\scriptscriptstyle 1\over
\scriptscriptstyle2}a_1(0)$ we can also have cases with
$a_1(q)={\scriptscriptstyle 1\over
\scriptscriptstyle2}a_1(0)$, $q> 1$,  and the spectrum is
\be {\rm Spect}(\hat A\, |\, q\neq 0)=\{(n-{\scriptstyle
1/2})a_1(0)\ |\, n=1, 2,
\cdots\
\}
\label{charged_spectrum}  
\ee for all these.  How do these conclusions square with
Barvinsky, Das and Kunstatter's~\cite{bar_das_kunst}
conclusion from canonical quantum gravity that $a_1(q)$ goes
up like $q^2$ ?  The apparent contradiction may reflect our
omission from the algebra of some operator, on the par with
$\hat A,\hat Q$ and $\hat {\bf J}$, which does not commute
with all of them.   Then some of the eigenvalues of $\hat A$
we have found with our sparser subalgebra may be ``killed'' by
consistency requirements.  In fact, this may be necessary
because according to Eq.~(\ref{area}), in the classical limit
$A\geq 4\pi Q^2$, so that area eigenvalues for large $q$ may
simply have to be large even for $n=1$.   

\section{The Degeneracy Factor and Black Hole Entropy}
\label{degen}

As mentioned in Sec.~\ref{Alg}, the area eigenvalue
corresponding to state $|nqjms\rangle$ may be degenerate to the
tune of $g_n$.  By the same symmetry arguments we applied in
Sec.~\ref{Spec} to conclude that $a_n=a_n(q)$ only, we conclude
that $g_n=g_n(j,q)$ only. 

Consider now the $g_{n_\kappa}$ different states $|\,n_\kappa,
j_\kappa, m_\kappa, q_\kappa,s\rangle$; we assume not both
$j_\kappa$ and $q_\kappa$ vanish.  And consider also the $g_1$
different states $\,|n_\lambda=1, j_\lambda=0, m_\lambda=0,
q_\lambda=0,s\rangle$.  We can form
$g_1(0,0)\cdot g_{n_\kappa}(j_\kappa,q_\kappa)$ states
$|\kappa,\lambda\rangle$ in the notation of Sec.~\ref{Spec},
which we shall suppose to be independent.  According to our
results in Sec.~\ref{Spec}, all the 
$|\kappa,\lambda\rangle$ have charge $q_\kappa$.  Further,
since we are combining zero angular momentum with $j_\kappa$,
it seems obvious (and can be established rigorously by use of
Eqs.~(\ref{RJz})-(\ref{RJpm})) that all the
$|\kappa,\lambda\rangle$  bear angular momentum $j_\kappa$. 
By Sec.~\ref{Spec} all our $|\kappa,\lambda\rangle$ here have
area quantum number $n_\kappa+1$.  Since there are a total of
$g_{n_\kappa+1}(j_\kappa,q_\kappa)$ one-black hole states with
these quantum numbers we must have
\be g_1(0,0)\cdot g_{n_\kappa}(j_\kappa,q_\kappa)\leq
g_{n_\kappa+1}(j_\kappa,q_\kappa).
\label{recursion}
\ee Henceforth we drop the subindex $\kappa$.

One immediate consequence of Eq.~(\ref{recursion}), given that
$g_1(0,0)\geq 1$, is that $g_{n+1}(j,q)\geq g_n(j,q)$:
degeneracy cannot decrease with $n$.  Let us iterate
(\ref{recursion}) starting from
$n_\kappa=1$ and assuming $g_1(0,0)\neq 1$:
\be g_n(j,q)\geq  g_1 (j,q)\cdot g_1(0,0)^{n-1}
\ee Thus $g_n(j,q)$ grows {\it at least\/} exponentially with
$n$.  We can think of
$\ln g_n(j,q)$ as the entropy associated with a black hole
with quantum numbers $j, m$ and $q$ simply because the
degeneracy represents a multiplicity of states which are
observationally indistinguishable.  Taking the logarithm and
comparing with
Eqs.~(\ref{neutral_spectrum})-(\ref{charged_spectrum}) shows
that the entropy must grow at least as fast as the area.  If
it grows exactly as area we get the usual black hole entropy
formula  $S\sim A$ in the classical limit.  The proportionality
factor, $\ln g_1(0,0)/a_1(0)$, must equal the
$(4\hbar)^{-1}$ of Hawking's formula. Adopting $g_1(0,0)=2$
for  illustration we deduce~\cite{bek_mukh95} 
\be a_1(0)=4\hbar\ln 2, 
\ee which calibrates the area spectrum
(\ref{neutral_spectrum})-(\ref{charged_spectrum}).  The
discrete mass spectrum (\ref{root_spectrum}) then follows from
Eq.~(\ref{mass}).

Another possibility is that entropy grows somewhat faster than
area.  For example, $g_n(j,q)\sim  n g_1(0,0)^{n-1}$.  In this
case we get a correction like $\ln A$ (with positive
coefficient) to the entropy.  This comes directly from the
degeneracy, not from quantum fluctuations, etc.

\vspace{0.5cm}

{\bf Acknowledgement:}  I thank Jim Hartle, John Schwartz and
Gilad Gour for incisive  remarks. This research is supported
by grant No. 129/00-1 of the Israel Science Foundation.

\end{document}